# The quadruple spectroheliograph of Meudon observatory (1909-1959)

J.-M. Malherbe, Observatoire de Paris, PSL Research University, Meudon, France

Email : Jean-Marie.Malherbe@obspm.fr                Date: 23 February 2024

ORCID identification : https://orcid.org/0000-0002-4180-3729

## ABSTRACT

The spectroheliograph was invented independently by Henri Deslandres (France) and George Hale (USA) in 1892, following the spectroscopic method suggested by Jules Janssen in 1869. This instrument is dedicated to the production of monochromatic images of the Sun in order to reveal the structures of the photosphere and the chromosphere at various altitudes. Sporadic observations started in Paris, but Deslandres moved soon to Meudon and designed, with Lucien d'Azambuja, an universal and powerful instrument, the quadruple spectroheliograph. It was devoted to systematic observations of the Sun (the long-term activity survey since 1908) and scientific research in solar physics. This paper describes the instrument and presents some original observations made with the high dispersion 7-metre spectrograph. It was dismantled in the sixties, but the solar patrol continued with the 3-metre chambers with Hα and CaII K lines, and is still working today with the numerical version of the spectroheliograph.

## 1 – INTRODUCTION

The principle of imaging spectroscopy of the solar surface was early introduced by Janssen (1869) after the discovery of how to observe prominences at any time, outside eclipses, using spectroscopic means: "with a spectroscope rotating around its axis, the entrance slit scans the Sun; a second slit in the spectrum selects a spectral line of interest and, using an ocular, the retinal persistence forms a monochromatic image". George Hale (USA) and Henri Deslandres (France) invented independently the spectroheliograph in 1892 (Malherbe, 2023) on this basis.

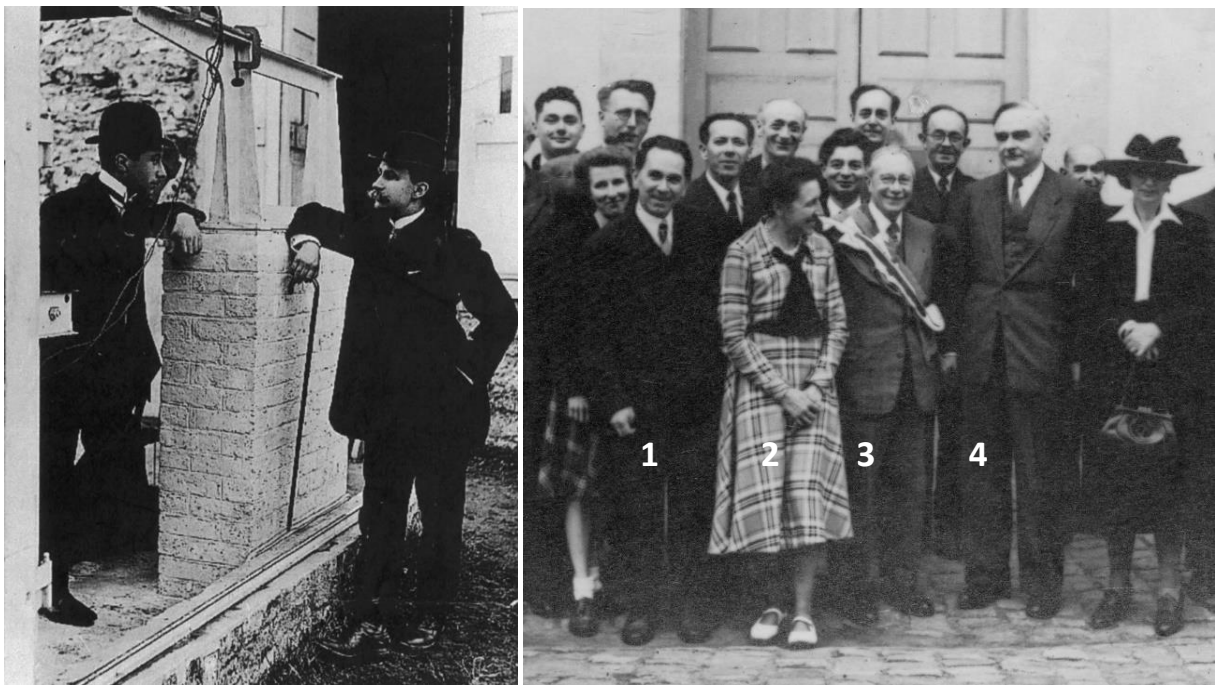

*Figure 1* : *Lucien d'Azambuja and Henri Deslandres (left), 1913. Lucien d'Azambuja's jubilee (50 years of astronomy) occured in 1949 and a ceremony was organized in 1950; among participants, (1) Bernard Lyot, (2) Marguerite d'Azambuja, (3) Lucien d'Azambuja, (4) André Danjon, director (courtesy OP).*

Deslandres (figure 1, see the biographies of d'Azambuja, 1948 and Lequeux, 2022), produced in Paris the first spectroheliograms of CaII K intensities using a thin slit in the spectrum integrating K2v, K3 and K2r (figure 2) and the first "section" spectroheliograms using a large slit (figure 3), which are made of the juxtaposition of CaII K spectra (cross sections by spatial steps of 30") for the measurement of radial velocities (see Deslandres, 1897).

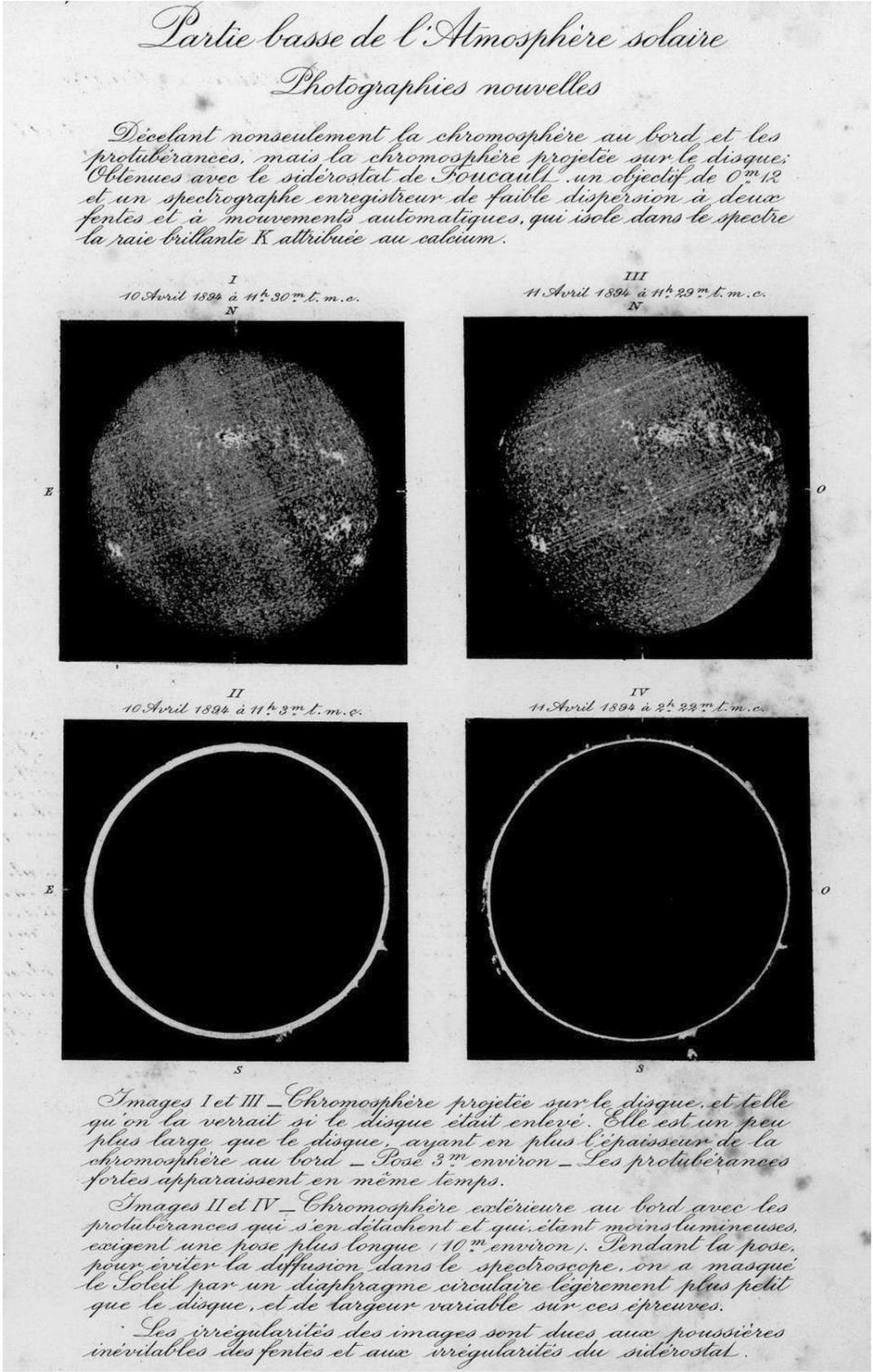

*Figure 2* : the first CaII K spectroheliograms of the disk ("chromosphère du disque", top) and of the limb ("chromosphère du bord", bottom) with a mask upon the disk. After Deslandres (1897). Courtesy CNUM/CNAM.

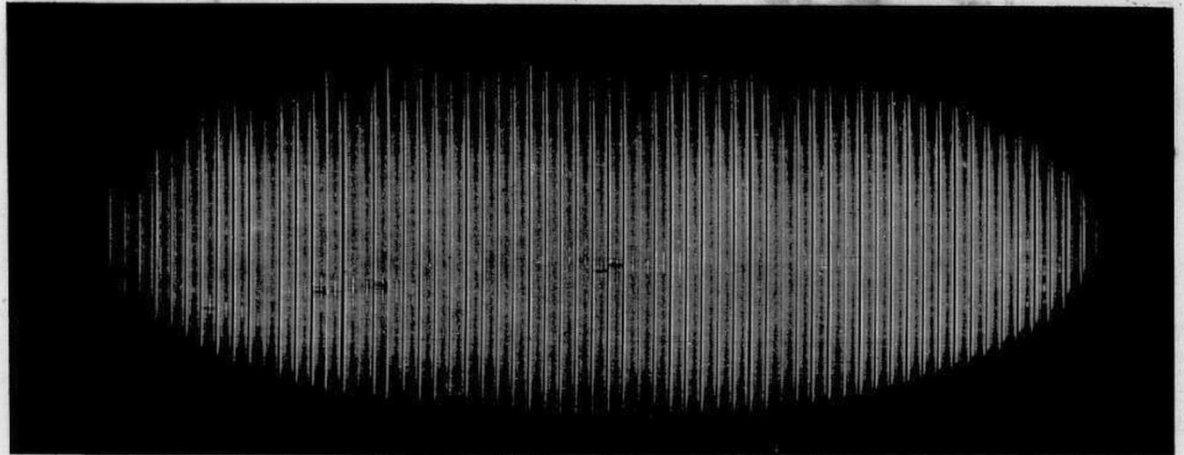

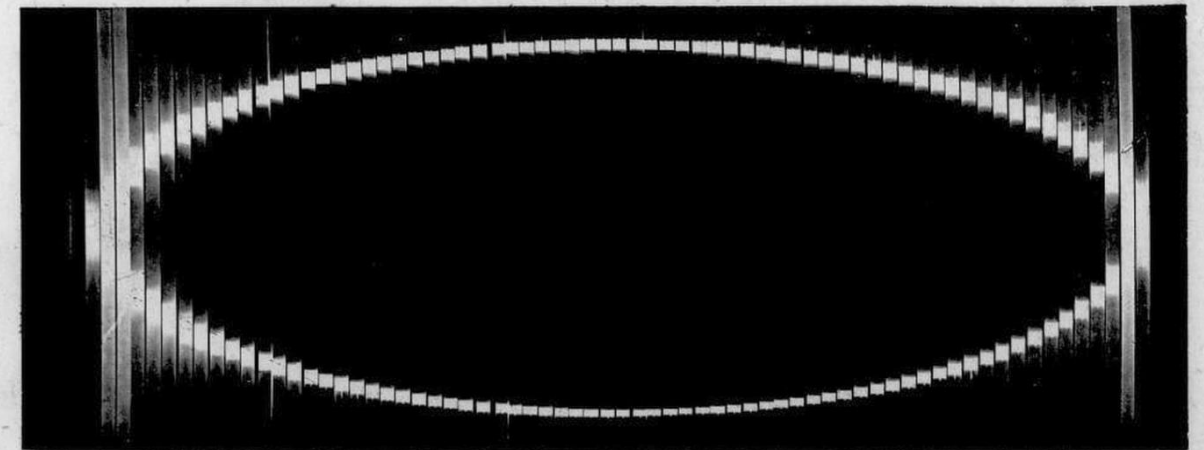

*Figure 3* : the first CaII K "cross section" spectroheliograms of the disk (top) and of the limb (bottom) with a mask upon the disk, showing CaII K spectra by spatial steps of 30". After Deslandres (1897). Courtesy CNUM/CNAM.

Deslandres moved to Meudon in 1898 and hired there an assistant, Lucien d'Azambuja (figure 1, see the biographies of Rösch, 1970 and Martres, 1998). They together built a much more powerful instrument than the one from Paris, the large quadruple spectroheliograph (section 2), which was completed in 1908. A removable spectrohelioscope was added later (section 3); systematic observations of the photosphere and chromosphere (section 4) started immediately, together with scientific research (section 5) leading to d'Azambuja's thesis (1930) and to the big memoir of d'Azambuga & d'Azambuja (1948) about the physics of solar filaments and prominences.

**2 – THE LARGE QUADRUPLE SPECTROHELIOGRAPH OF MEUDON**

The quadruple spectroheliograph was built by Deslandres (1910) and d'Azambuja in a new house (1907) and was fed by a two mirror coelostat (figure 4). Four chambers were installed in a 20-metre long laboratory. Chambers n° I and n° II had moderate dispersion and were devoted to systematic observations, respectively with CaII K and Hα lines. Chambers n° III and n° IV were dedicated to scientific research, for multi-line or high dispersion observations. They could not work simultaneously. The entrance objective (250 mm diameter, 4 m focal length), entrance slit, collimator, grating and prisms were common to all combinations, which are detailed in figures 5.

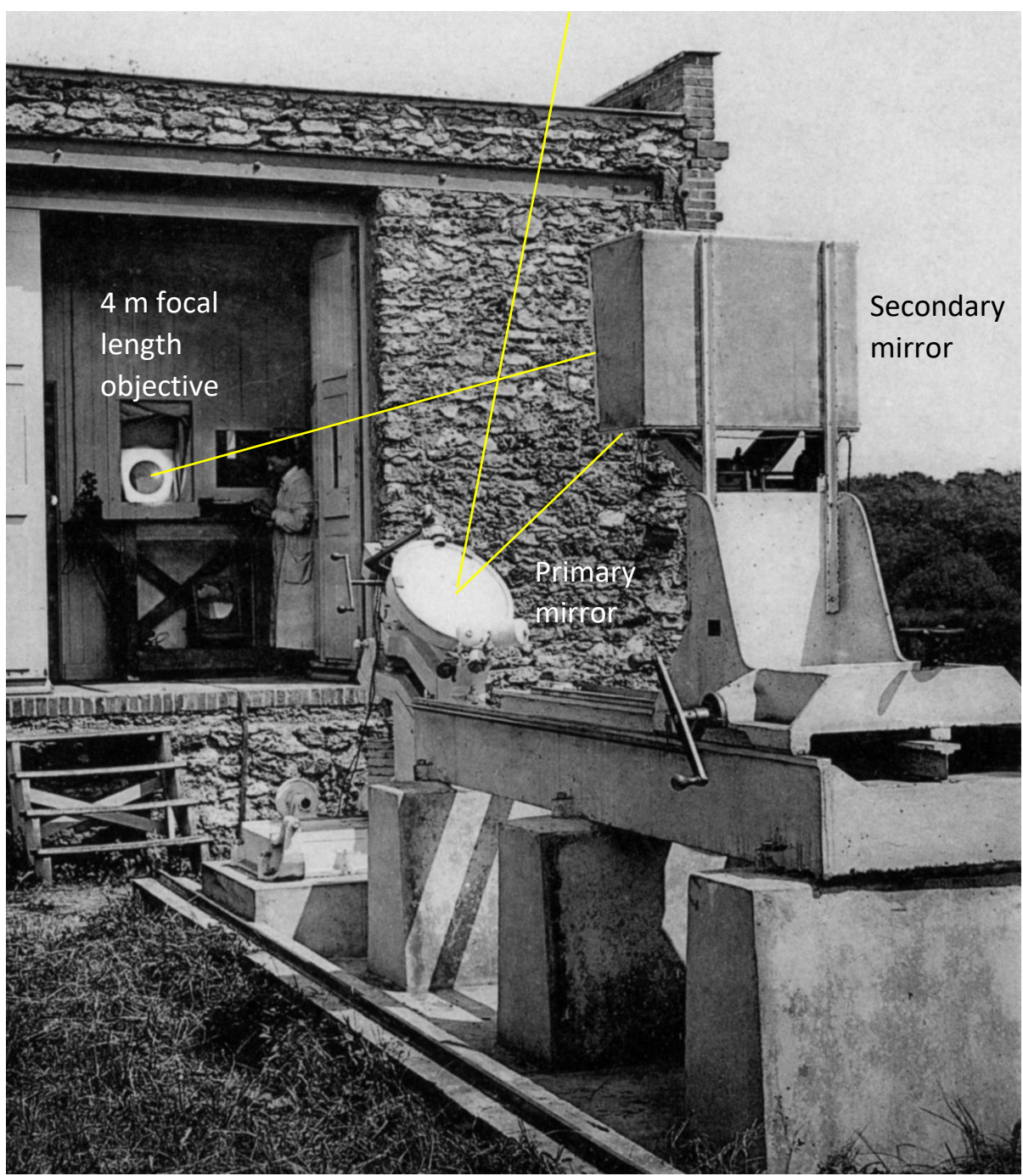

*Figure 4 : the two-mirror coelostat feeding the quadruple spectroheliograph (courtesy OP).*

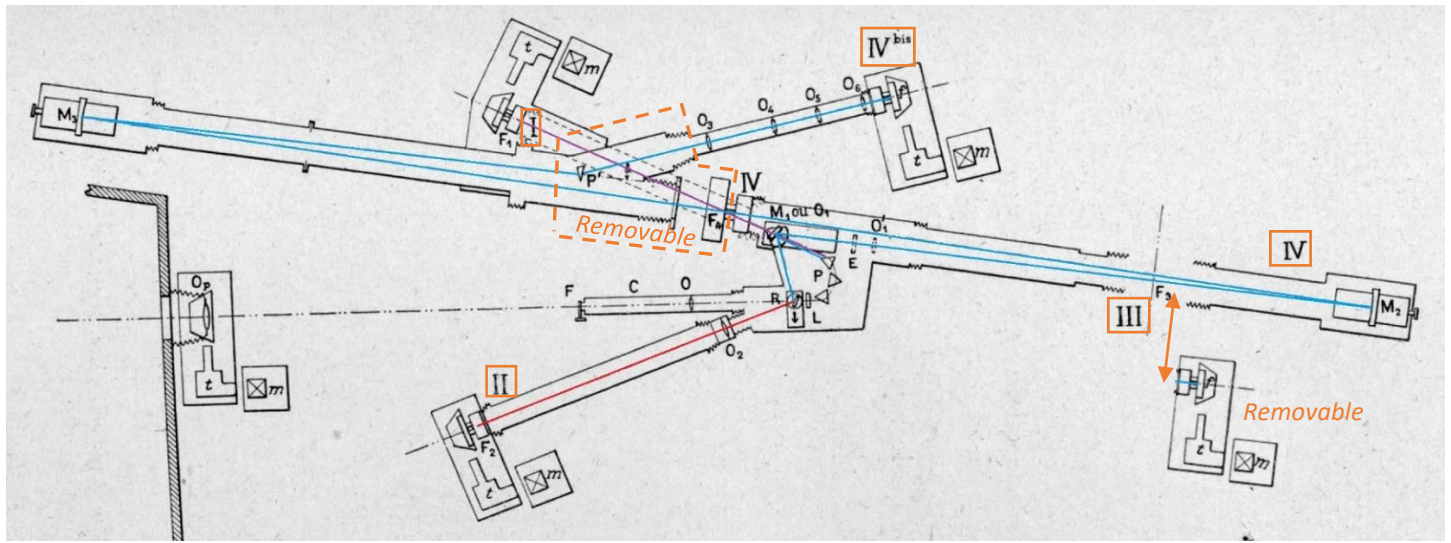

*Common parts* : Op (4 m objective), F (entrance slit), O (collimator, f = 1.3 m), R (grating) or P (3 prisms, flint, 60°)

*Spectrograph/chamber n° I (CaII K): 86 mm image*

    *P (3 prisms), $O_1$ (chamber, f = 3.0 m), $F_1$ (output slit), t, m (motorized plate carrier)*

*Spectrograph/chamber n° II (Hα): 86 mm image*

    *R (grating), $O_2$ (chamber, f = 3.0 m), $F_2$ (output slit), t, m (motorized plate carrier)*

*Spectrograph/chamber n° III (multi-line, multi-purpose): 86 mm image*

    *P or R, $M_1$ (flat mirror), $O'_1$ (chamber, f = 3.0 m), $F_3$ (output slit), t, m (motorized plate carrier)*

*Spectrograph/chamber n° IV (multi-line, high dispersion): 205 mm image*

    *P or R, $M_1$ (flat mirror), $M_2$ (chamber, f = 7.0 m), $F_4$ (output slit)*

    *Associated afocal reducer n° IVbis, low dispersion: various image size, according to the reduction factor*

    *$M_3$ (collimator, f = 7.0 m), P' (prism), $O_3/O_4/O_5/O_6$ (chamber), f (output slit), t, m (motorized plate carrier); reduction factor = focal length ratio $M_3/O_{3,4,5,6}$ (4 possible combinations)*

*Removable parts: t,m of chamber III, $F_4$ (spectrograph IV), P' (disperser IVbis), tubes of chamber I and reducer IVbis*

*Motorized translators: t,m for entrance objective Op and photographic plate carriers*

*Manual translators: R or P, $M_1$ (for chamber IV) or $O_1$ (for chamber III)*

> **Figure 5** : optical parts of the Meudon quadruple spectroheliograph (after Deslandres, 1910, and d'Azambuja, 1930). Some of the different combinations were not simultaneously available (courtesy OP).

The chamber n° II was used for systematic observations of Hα with a plane grating (figure 6), while chamber n° I was dedicated to CaII K with a set of three prisms (figure 6). Both chambers were compatible with chamber n° III which was there for research purpose, allowing multi-line observations, with similar dispersion provided by the same focal length (spectra of 86 mm height, formed by the $O_1$, $O_2$, and $O'_1$ lenses, respectively for chambers n° I, II, III). The beam was directed towards the spectrograph n° III by the flat mirror $M_1$. The use of the high dispersion spectrograph n° IV was more complicated, because some parts had to be introduced, and other removed. As the focal length of the concave mirror $M_2$ was 7 m, the height of the spectrum at the slit $F_4$ was 205 mm. An associated afocal reducer IVbis, composed of the combination of a second 7 m concave mirror $M_3$ and one of the four lenses $O_3$, $O_4$, $O_5$, or $O_6$, allowed to reduce the image size without any alteration of the spectral resolution determined by the width of the slit $F_4$. P' was a 30° prism, it was there to eliminate parasitic light (or orders) from the grating. The support of $F_4$ and the tube of the reducer IVbis had to be dismounted to permit daily observations with the CaII K chamber n° II. The dispersion provided by the four spectrographs is given by figure 7. It was possible to attain about 1 Å/mm for scientific research with chamber n° IV from the near Ultra Violet to the near Infra Red.

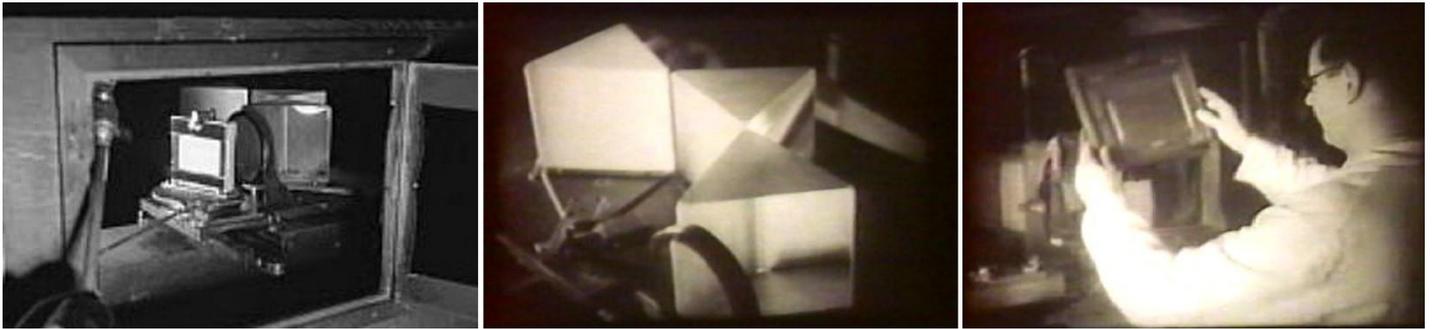

*Figure 6* : the plane grating (left), the set of three prisms (centre) and the motorized photographic plate carrier (right). Courtesy OP.

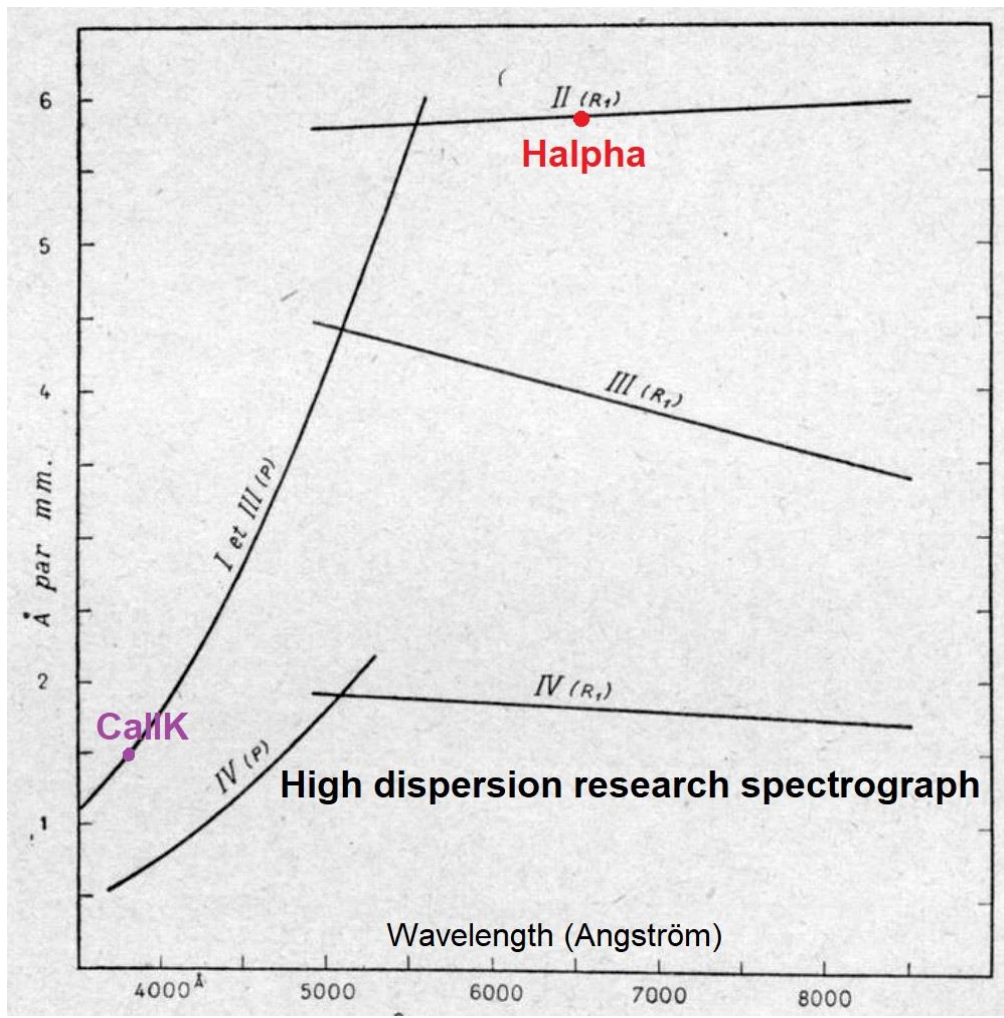

*Figure 7* : the dispersion of the spectrographs in Å/mm. Chamber n° I (prisms) and n° II (grating) were used for systematic observations of CaII K and Hα. The spectrograph n° IV, with 7 m focal length instead of 3 m, provided the best dispersion and was dedicated to solar research. Courtesy OP, after d'Azambuja (1930).

Figure 8 shows the quadruple spectroheliograph configured for the use of the high dispersion chamber n° IV (7 m focal length, spectrum of 205 mm height). The associated afocal reducer n° IVbis was used to reduce the height of the spectrum in order to be more compatible with classical photographic plates (13 x 18 cm² max). The reduction factor was equal to the ratio between the focal length of the concave mirror $M_3$ (f = 7 m) and the focal length of one of the available lenses $O_3$, $O_4$, $O_5$ or $O_6$. The reducer included a low dispersion 30° prism to eject parasitic light. The dispersion of the chamber n° IV was convenient to produce monochromatic images with a very narrow bandpass (as low as 0.05 Å). For that reason, it was employed for scientific purpose, not for daily observations.

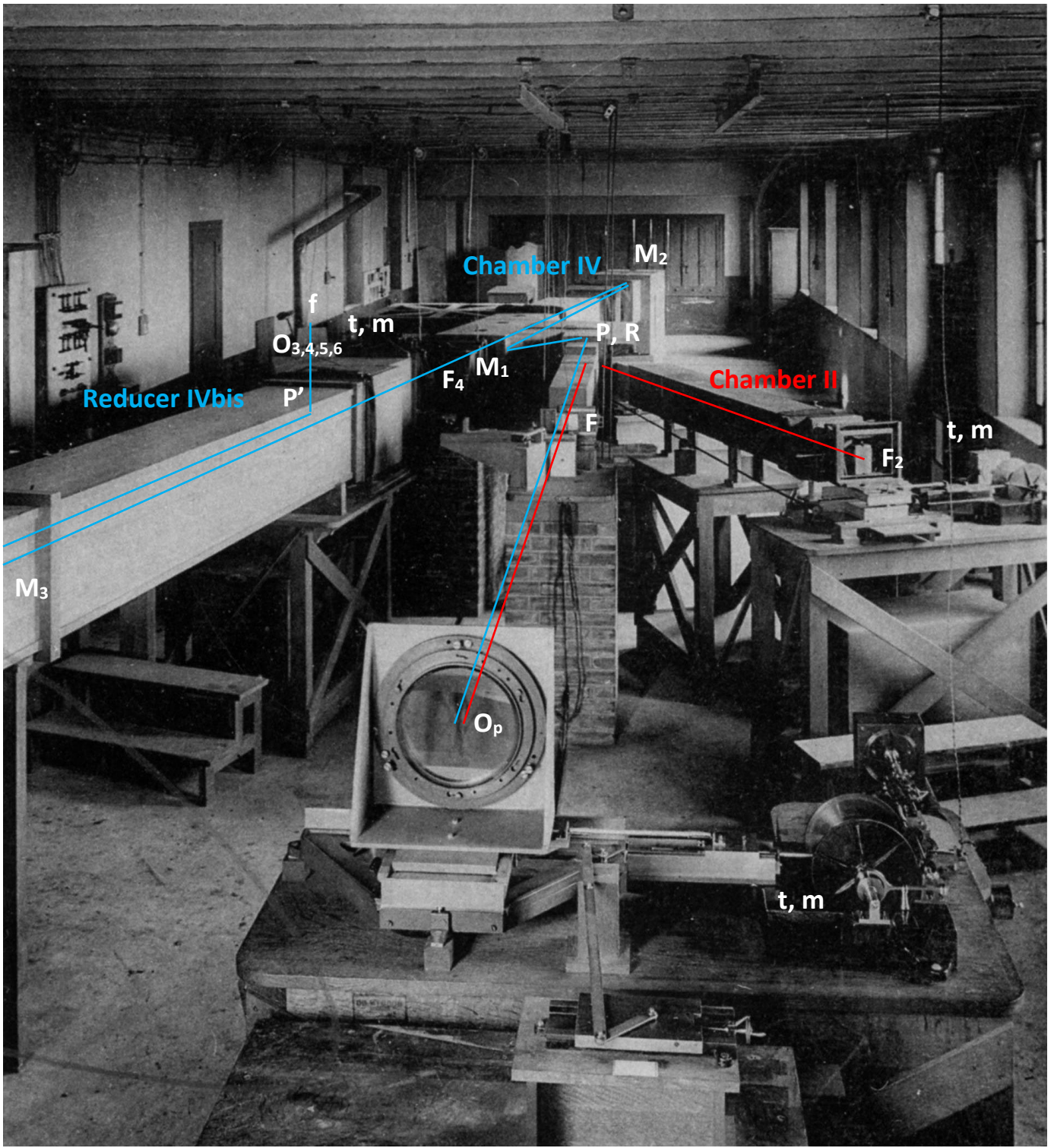

***Figure 8*** : *Meudon quadruple spectroheliograph configured for the use of the high spectral resolution chamber n° IV (7 m focal length) and associated focal reducer n° IVbis (the letters refer to figure 5). The moderate resolution chamber II for systematic observations of Hα remained available (courtesy OP).*

Alternatively, Figure 9 shows the quadruple spectroheliograph configured for systematic observations of the photosphere and the chromosphere with chambers n° I (CaII K) and n° II (Hα). For that purpose, the slit $F_4$ of chamber n° IV had to be displaced, together with several parts of the associated reducer n° IVbis which was used in combination with it. CaII K (K1v for the photosphere and K3 for the chromosphere) was observed with the set of three prisms, while Hα was formed by the plane grating. The instrument and observing program are described by d'Azambuja (1920 a and 1920 b).

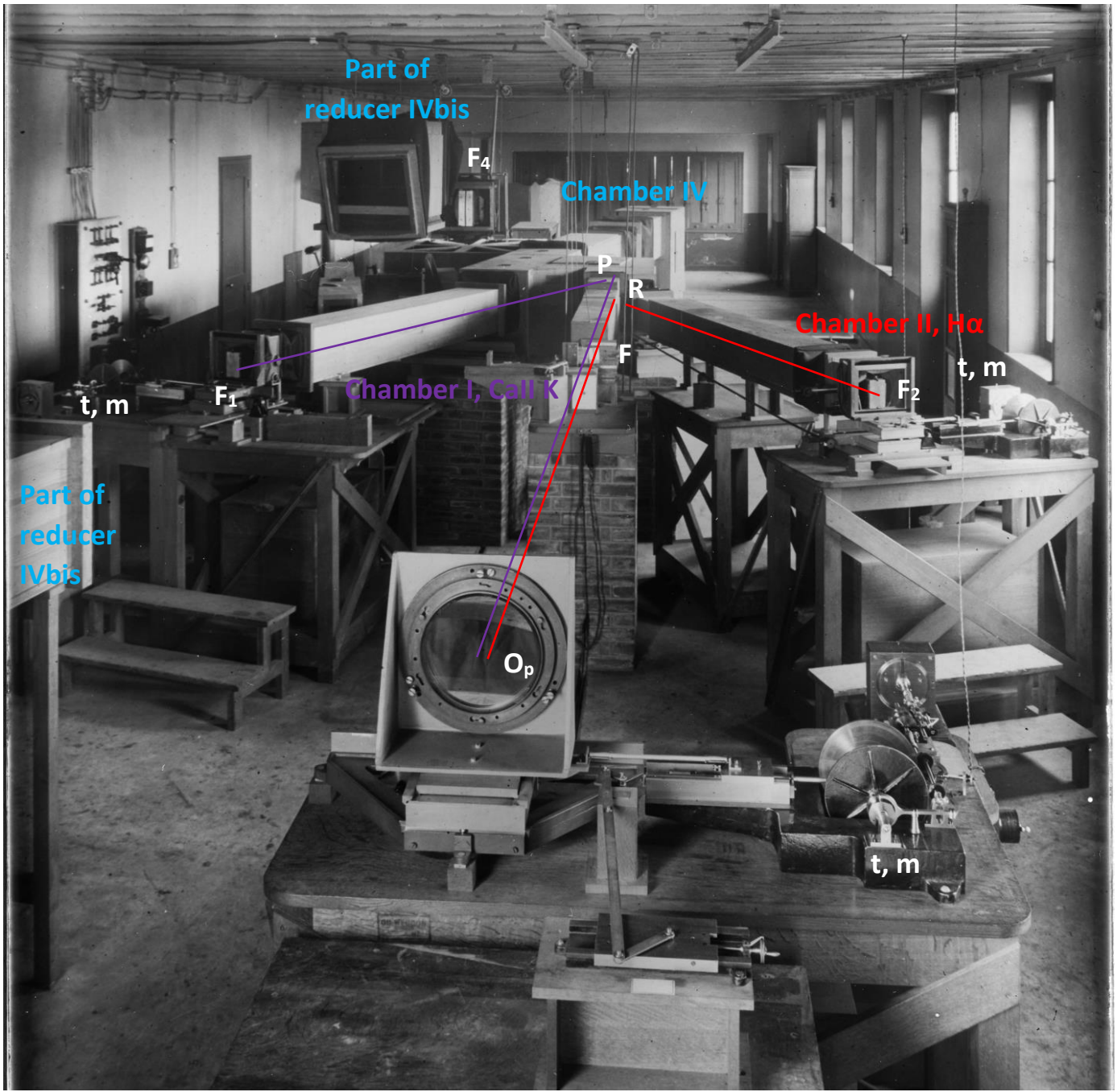

*Figure 9* : *Meudon quadruple spectroheliograph configured for systematic observations of CaII K and Hα with moderate resolution chambers n° I and n° II (courtesy OP). The chamber n° IV and associated afocal reducer n° IVbis were partly dismounted for daily observations (the letters refer to figure 5, courtesy OP).*

## 3 – THE VISUAL SPECTROHELIOSCOPE

Figure 10 shows the spectroheliosocope which was installed later (1940) by d'Azambuja. It was a visual device, using two Anderson's square prisms (figure 11). The first prism, in front of the entrance slit F of the spectrograph, rotated in order to scan active regions of the Sun. The second prism, rotating at the same speed, was located after the slit $F_2$ which was isolating the Hα line in the spectrum. The size of the square prisms were in the ratio (2.3) of the focal lengths of the collimator (1.30 m) and the chamber (3.0 m). The observer used an ocular (figure 11) to see the monochromatic image formed by the output slit $F_2$ and the retinal persistence. This system was very fast to install and remove (to allow photographic observations), and was of great interest for visual inspection of solar activity, so that the observers used it daily.

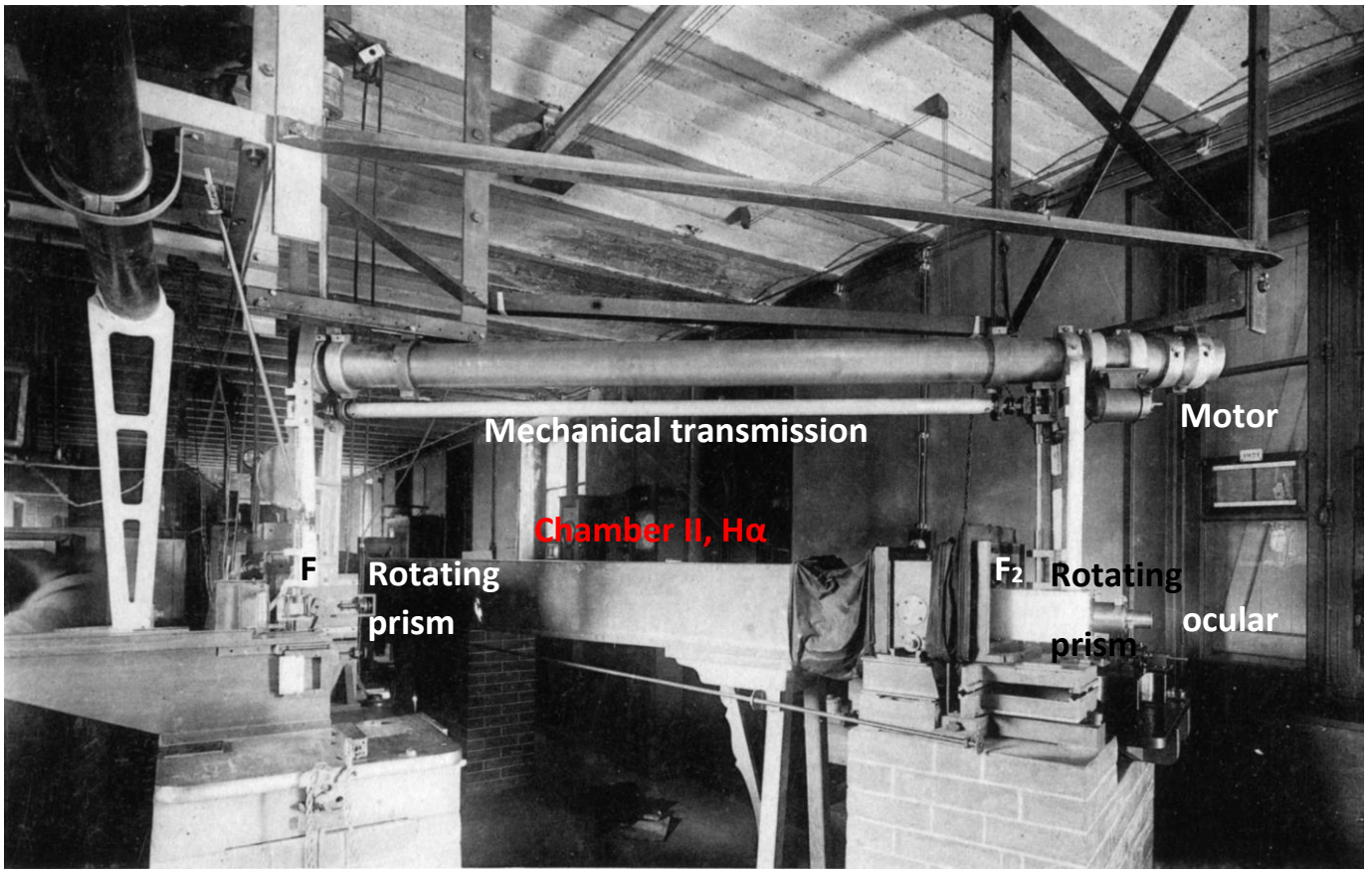

*Figure 10* : *Meudon visual spectrohelioscope for Hα using two high speed rotating prisms and the moderate resolution chamber n° II (courtesy OP).*

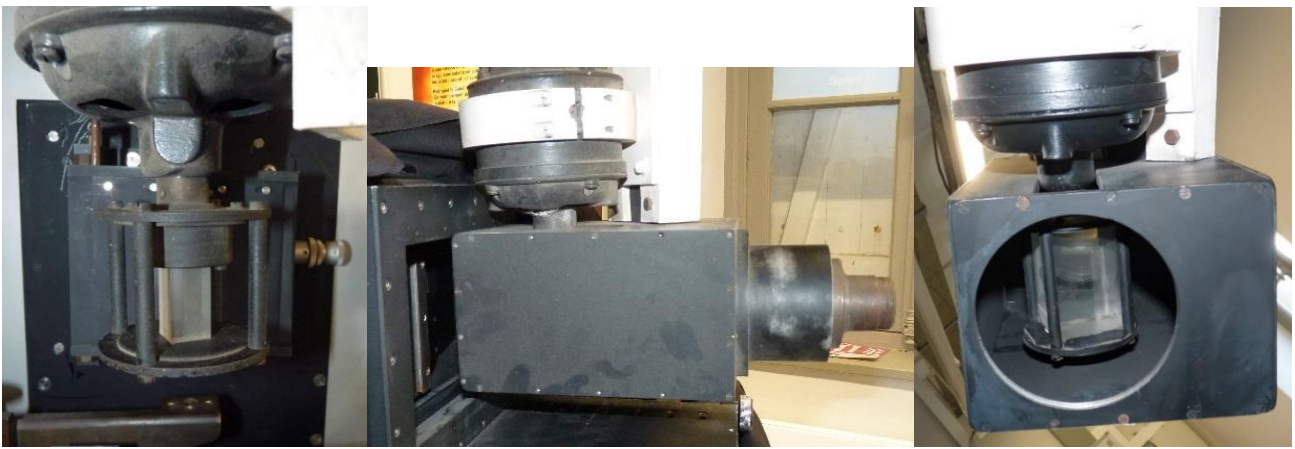

*Figure 11* : *the rotating prism in front of the entrance slit F of the spectrograph (left), the ocular (centre) and the rotating prism at the output slit $F_2$ (right). Courtesy OP.*

### 4 – SYSTEMATIC OBSERVATIONS WITH CHAMBERS n° I AND n° II

Systematic observations were performed with the configuration of figure 9: spectrograph n° I for CaII K using 3 flint prisms of 60° at minimum deviation, providing a spectral resolution of 0.15 Å, and spectrograph n° II for Hα using a Rowland plane grating, 568 groves/mm, providing in the first order a spectral resolution of about 0.40 Å. The magnification of the spectrographs was 2.31 (3.0 m chambers, 1.30 m collimator), so that the size of the Sun on the photographic glass plates was 86 mm. The width of the entrance slit was 0.03-0.04 mm (1.5"-2") well adapted to the usual seeing of 2" in Meudon; while the second slit in the spectrum was 0.075 mm wide.

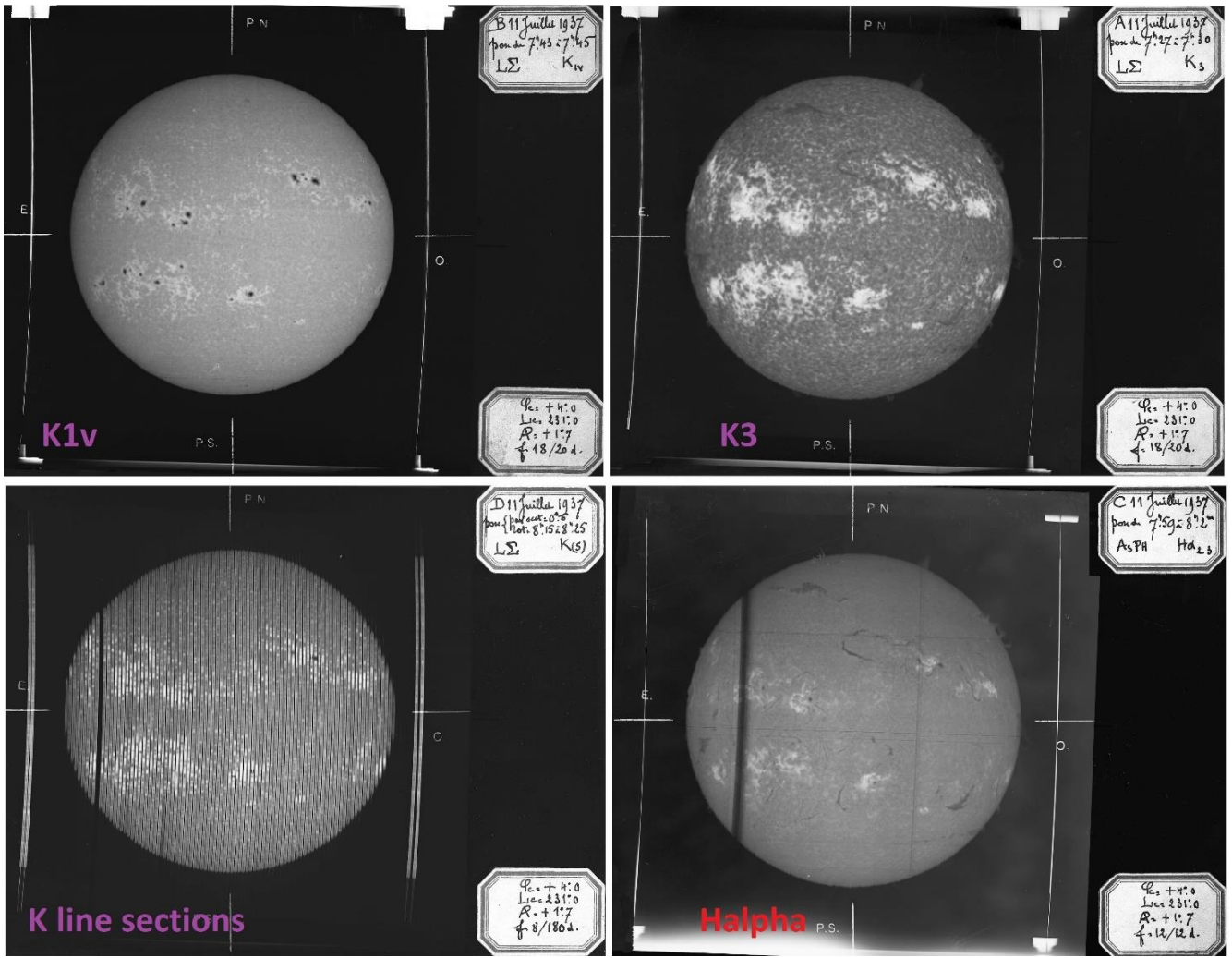

*Figure 12. Typical spectroheliograms of the period 1919-1939 (example of 11 July 1937). CaII K1v for the photosphere (sunspots, faculae); CaII K3 for the chromosphere (filaments, bright plages); CaII K line profiles (cross sections of the Sun by 22" steps for Dopplershift measurements); Hα for the chromosphere (filaments, plages). Courtesy OP.*

From 1908 to WW1 (1914), observations consisted of daily spectroheliograms in CaII K3 (line centre, chromosphere), CaII K1v (violet line wing, photosphere) and Hα centre (chromosphere). Observations were totally interrupted during WW1. Figure 12 shows typical observations performed daily from 1919 to 1939 until the start of WW2. During this period, CaII K3, CaII K1v and Hα continued, and a "section" spectroheliogram with an enlarged slit in the spectrum (1 mm = 2 Å) was added for Dopplershift measurements of the CaII K line; for that purpose, full spectra of 86 cross sections of the solar surface were recorded by spatial steps of about 1 mm = 22". Contrarily to WW1, observations were not stopped by WW2, but the lack of manpower was an important constraint, so that "section" spectroheliograms were abandoned, and never restarted after WW2. Typical systematic observations done after WW2 include the initial choice of 1909 (CaII K3, K1v and Hα), plus a long exposure CaII K3 for limb features and prominences, with a disk attenuator (because prominences are 5-10 times fainter than the disk).

Systematic observations continued under the direction of Lucien d'Azambuja until his retirement in 1954 (he was 70 years old). At this date, Mrs d'Azambuja became in charge of observations, but his husband was still working as a volunteer; she retired in 1959. Gualtiero Olivieri (1928-2018) and Marie-Josèphe Martres (1924-2014) became respectively responsible of observations and of the synoptic map program initiated in 1919 by L. d'Azambuja (these are synthetic drawings of solar activity for each solar rotation, based on daily spectroheliograms, and showing the average position of filaments, sunspots and facular regions).

## 5 - SPECIAL SPECTROHELIOGRAMS WITH THE HIGH DISPERSION CHAMBER n° IV

During his thesis work, that he defended in 1930, Lucien d'Azambuja explored many spectral lines with the large 7 m spectroheliograph devoted to research (figure 8). High dispersion is required to study thin photospheric lines such as those chosen by d'Azambuja: MgI 3838 Å, SrII 4078 Å, FeI 4202 Å, CaI 4227 Å, FeI 4384 Å, MgI 5184 Å, HeI 5876 Å, NaI 5890 Å, as well as chromospheric infrared lines of CaII 8498 Å, CaII 8542 Å, HeI 10830 Å. This large spectroheliograph (dismantled in the sixties) was able to offer the dispersion of 1 mm/Å, providing excellent spectral resolutions on line profiles of about 0.05 Å. The set of 3 prisms was used in the blue part of the spectrum, and the plane grating for longer wavelengths.

Figure 13 displays three images, respectively in the centre of Hα line and in the blue wing. Filaments are still visible at -0.25 Å, but vanish at -0.50 Å (near the inflexion points of the line profile) and the photosphere (below the chromosphere) begins to appear at -0.50 Å, except in the case of Dopplershifts (0.5 Å = 25 km/s). Such observations are of interest, because they were never done daily, contrarily to the Hα centre.

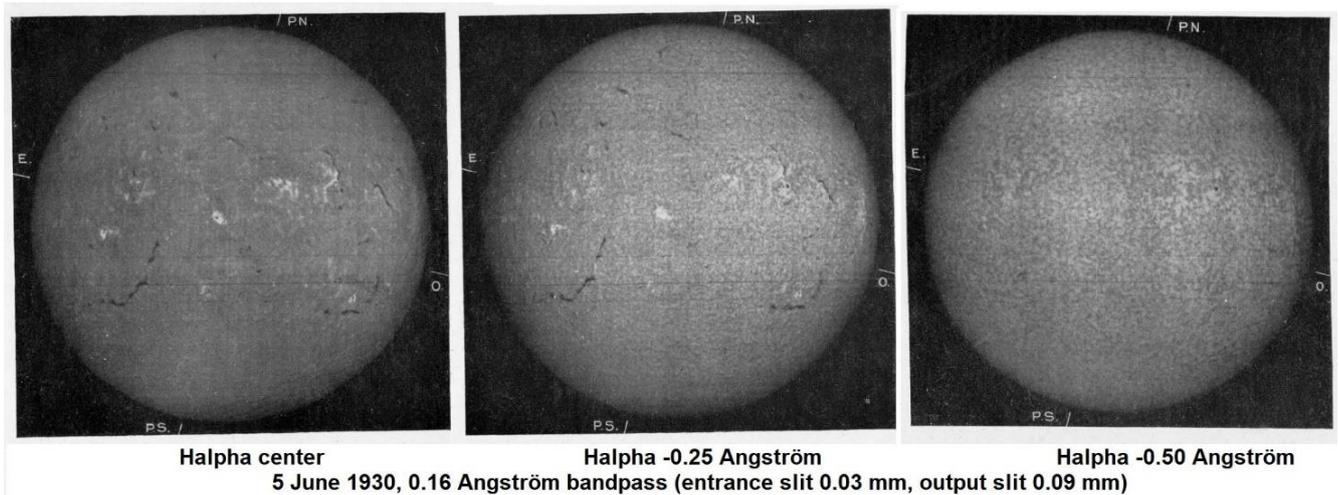

*Figure 13. Observations at three positions in the Hα line. Filaments vanish in the line wing, except in the case of Dopplershifts, while photospheric structures begin to appear. 5 June 1930. After d'Azambuja (1930).*

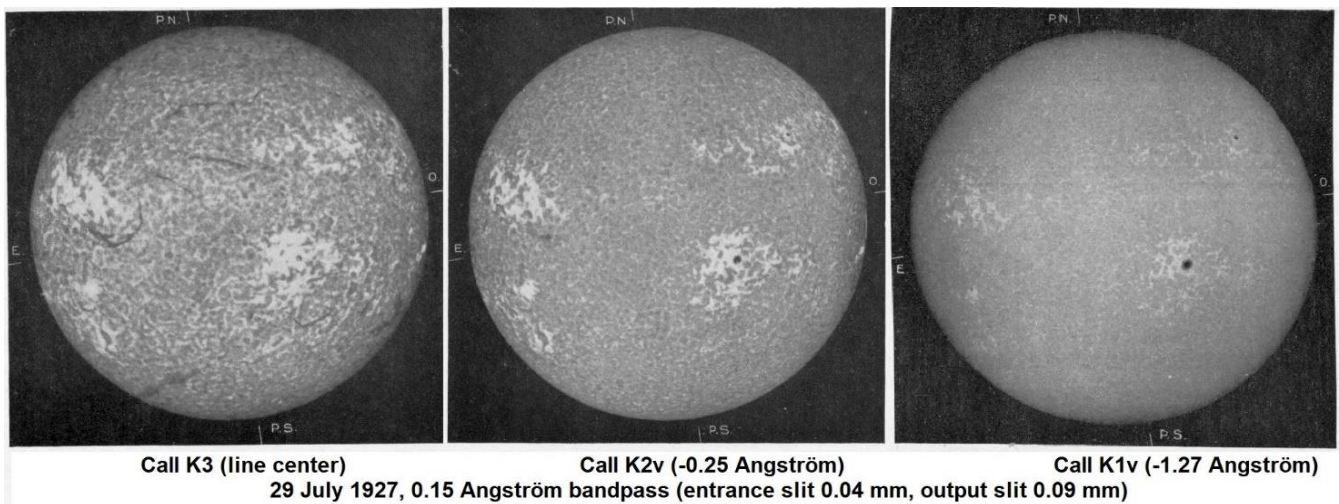

*Figure 14. Observations at three positions in the CaII K line (K3, K2v, K1v). 29 July 1927. After d'Azambuja (1930). Courtesy OP.*

Figure 14 displays three images, respectively in the centre of CaII K line (K3) and in the blue wing. Filaments and bright plages are perfectly visible in K3, but this is not the case of sunspots. Filaments are fairly observable at -0.25 Å (K2v) and sunspots begin to appear. But filaments completely vanish at -1.27 Å (K1v) while the photosphere, below the chromosphere, becomes clearly visible, with sunspots and bright faculae (which are associated to chromospheric plages). Observations of K2v are particularly rare in the collection, contrarily to K3 and K1v which are systematic.

Figure 15 shows an exceptional observation of the Balmer series, Hα, Hβ, Hγ, Hδ and Hε. The contrast of filaments and plages (chromosphere) vanishes with decreasing wavelength, because lines form deeper. On the contrary, sunspots appear progressively along the series, which reveals that Hα is the best choice for chromospheric structures, while Hδ or Hε are not better than CaII K1v for sunspots and faculae (however Hε is contaminated by the red wing H1r of the CaII H line). This observational series is the only one of the Meudon collection.

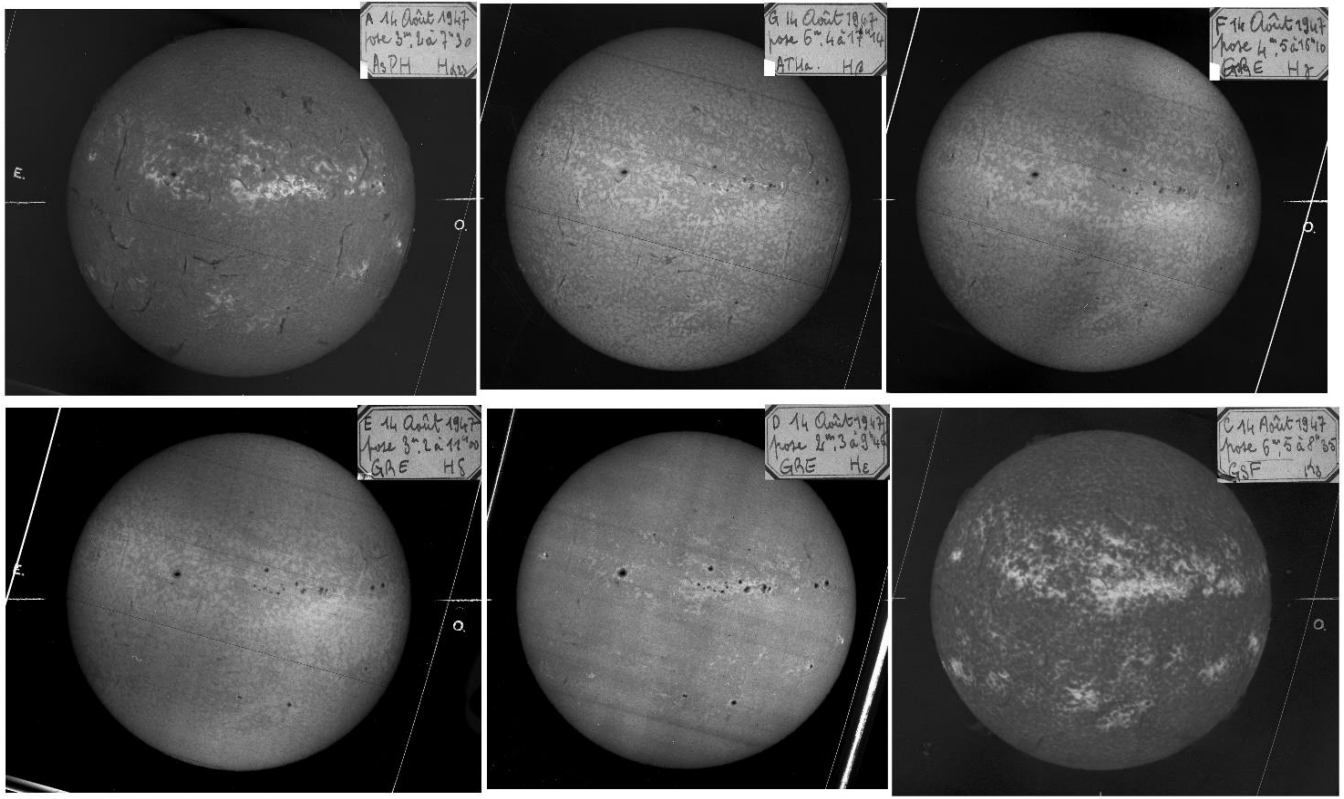

*Figure 15. The Balmer series Hα, Hβ, Hγ, Hδ and Hε of Hydrogen, 14 August 1947, and the CaII K3 spectroheliogram for comparison. Hα is chromospheric, but Hδ and Hε are rather photospheric. Courtesy OP.*

Helium lines where also studied by Mr and Mrs d'Azambuja. Figure 16 shows an exceptional observation in HeI D3 5876 Å of an active region producing the huge flare of 26 July 1946, while figure 17 shows the first world-wide observation done in the infrared HeI 10830 Å line. These observations are absolutely unique in the collection (d'Azambuja & d'Azambuja, 1938). Several decades later, Kitt Peak National Observatory started systematic observations in this infrared line of neutral Helium.

Infrared lines of CaII where investigated by d'Azambuja (1930) in his thesis work. Figure 18 shows images in CaII 8498 Å and 8542 Å, on 4 September 1928, in comparison with the usual CaII K3 3934 Å. These observations are extremely rare in Meudon collection and showed that chromospheric structures are better seen with CaII K3 (figure 19). Several decades later, Kitt Peak National Observatory started systematic observations in CaII 8542 Å in order to measure chromospheric magnetic fields (the Zeeman sensitivity is much better for this infrared line than for violet lines, such as CaII K or CaII H).

Figure 19 shows also a detail of figure 18, for comparison of HeI 10830 Å and CaII K. While filaments appear dark in both lines, this is not the case of plages, which are bright in CaII K but dark in HeI 10830 Å.

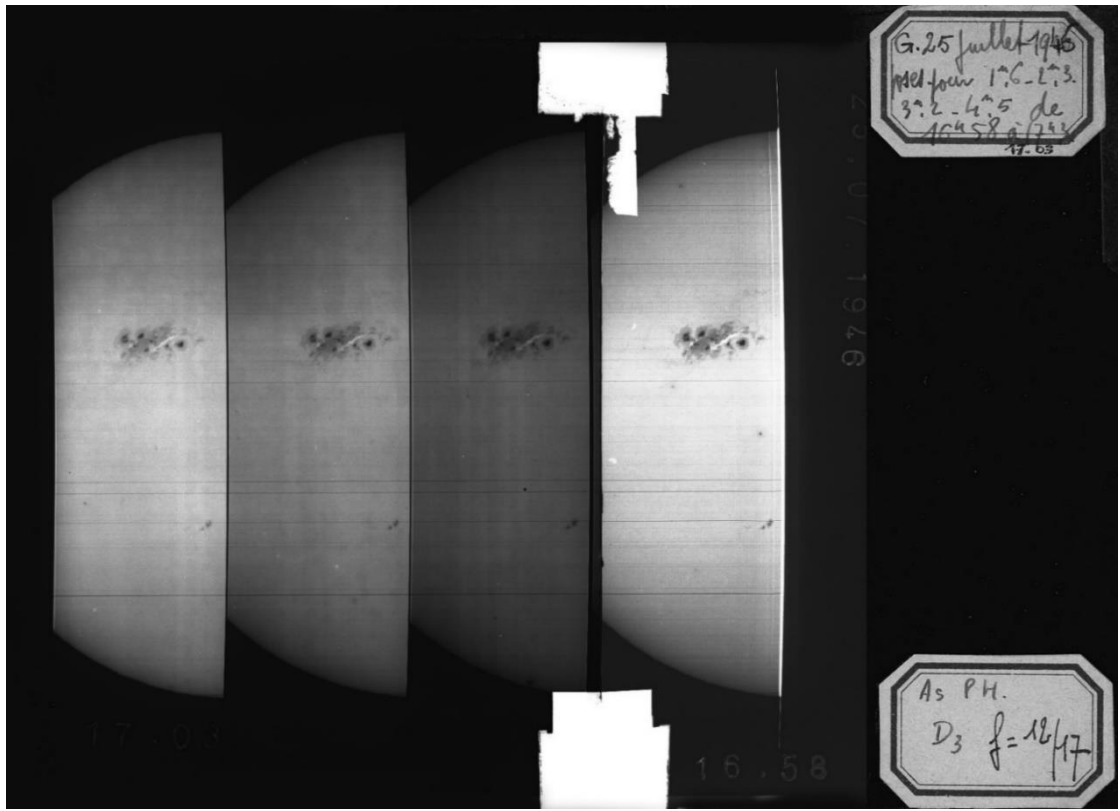

*Figure 16. The flare of 25 July 1946 observed in HeI D3 at 5876 Å from 16:58 to 17:03. Courtesy OP.*

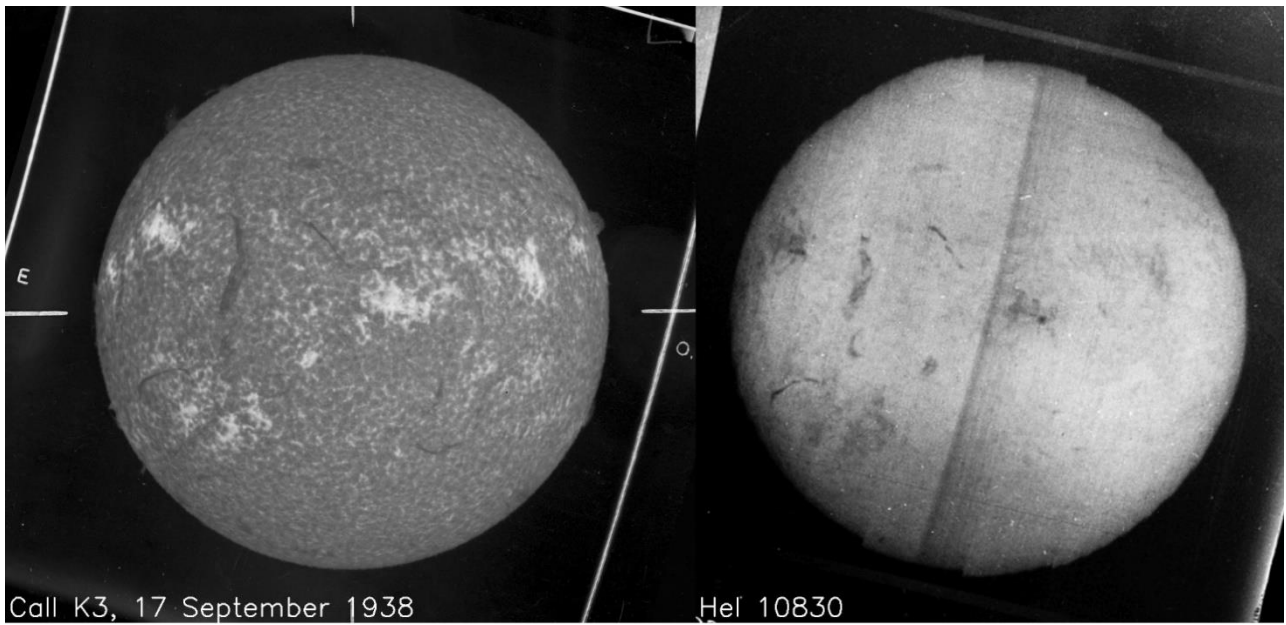

*Figure 17. The first observation of the infrared HeI 10830 Å line (right), where both filaments and plages appear darker than the quiet Sun. CaII K3 (left) is given for comparison. 17 September 1938. Courtesy OP.*

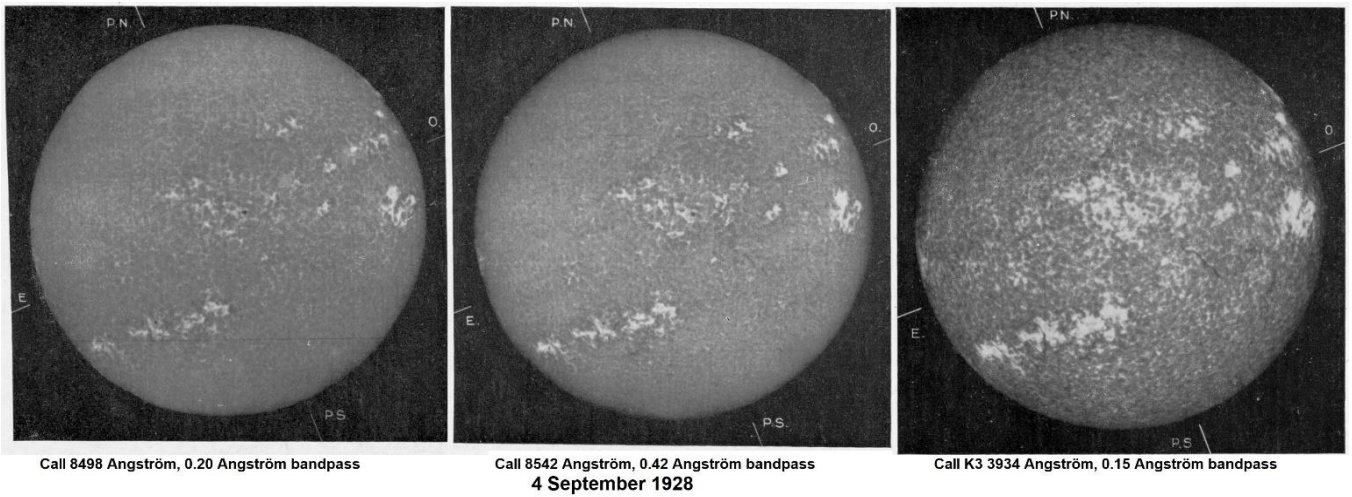

*Figure 18. CaII 8498 Å and CaII 8542 Å, compared to CaII K3 3934 Å (right). 4 September 1928. After d'Azambuja (1930) and courtesy OP.*

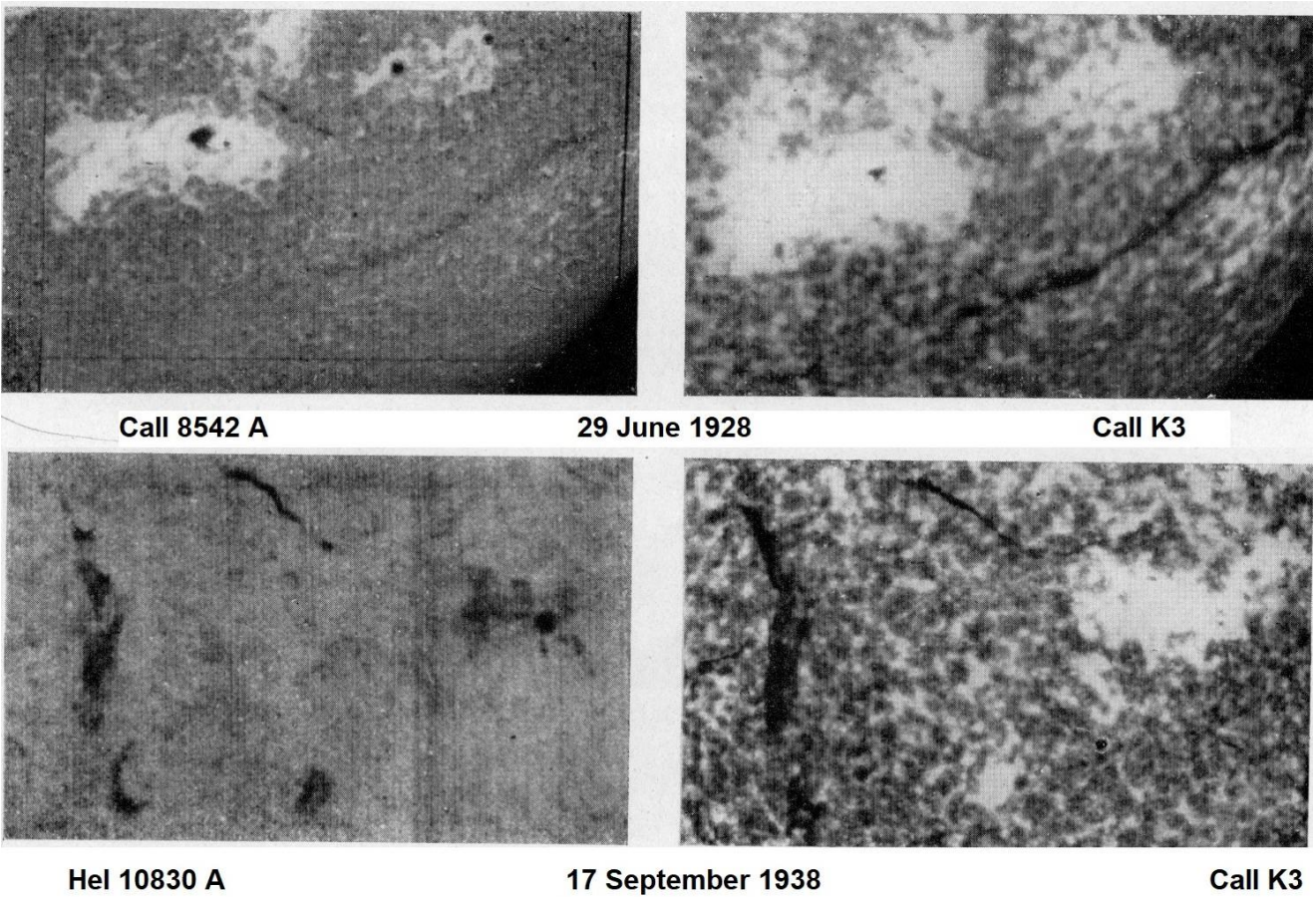

*Figure 19. CaII 8542 Å, compared to CaII K3 3934 Å (top). 29 June 1928. After d'Azambuja (1930). HeI 10830 Å, compared to CaII K3 3934 Å (bottom). 17 September 1938. Courtesy OP.*

CaI 4227 Å line, as well as many other photospheric lines, was studied by d'Azambuja (1930) during the course of his thesis. Figure 20 shows images in the line core and wing, in comparison with the usual CaII K1v spectroheliogram (the blue wing of CaII K line) for the photosphere. It sounded clearly that this Calcium line (and many thin lines) provided spectroheliograms that are comparable to CaII K1v, so that observations in these explored wavelengths were never reproduced later.

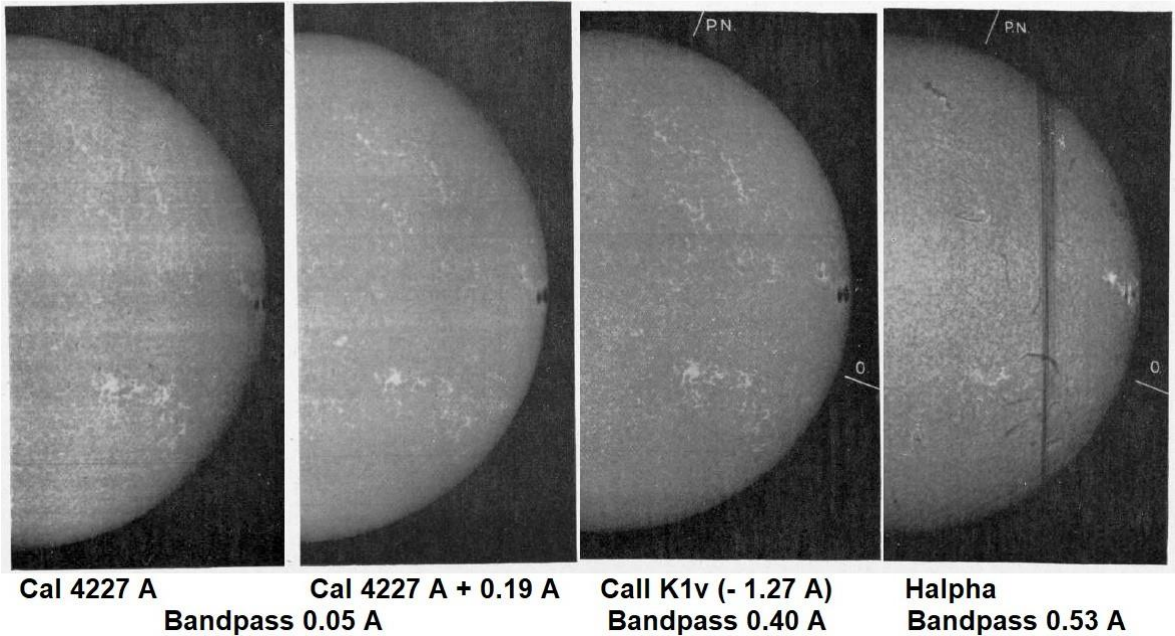

*Figure 20. Spectroheliograms of the CaI 4227 Å line, compared to CaII K1v for the photosphere (at right, for reference, the chromospheric Hα line). After d'Azambuja (1930) and courtesy OP.*

## 6 – CONCLUSION

The spectrographs n° III, n° IV and associated reducer n° IVbis remained unique during many decades and were dismantled after the retirement of Lucien and Marguerite d'Azambuja. However, parts of the 7 m spectrograph n° IV were transformed in the same building after 1960 for the measurements of magnetic fields through the Zeeman effect, using a new telescope (a 40 cm Newtonian telescope fed by a 80 cm Foucault siderostat, independent of systematic observations). The Zeeman analyser was the Hale's grid, composed of alternatively quarter (λ/4) and three quarter (3λ/4) wavelength strips (of a few arc seconds wide) and a polaroid (transmitting σ+ and σ- Zeeman components). λ/4 and 3λ/4 measurements were simultaneous, but not co-spatial. This was the starting point of many developments of solar magnetometry techniques which will be related in another paper. Systematic observations in CaII K, CaII H and Hα are still active, with a modernized version of chamber n° II, but now use a fast scientific CMOS detector.

## 7 - REFERENCES